\documentclass[fleqn,usenatbib]{mnras}

\usepackage{mathptmx}

\usepackage[T1]{fontenc}
\usepackage{ae,aecompl}

\usepackage{graphicx}	
\usepackage{amsmath}	
\usepackage{amssymb}	

\newcommand{\beq}{\begin{eqnarray}}
\newcommand{\eeq}{\end{eqnarray}}

\newcommand{\teq}{$t_{\rm eq}$}

\title[GW170817 Kilonova afterglow]{Observable Features of GW170817 Kilonova Afterglow}

\author[Kathirgamaraju, Giannios \& Beniamini]{Adithan Kathirgamaraju$^{1}$\thanks{E-mail: akathirg@purdue.edu (AK)}\thanks{Bilsland Fellow}, Dimitrios Giannios$^{1}$ and Paz Beniamini$^{2}$
\\
$^{1}$Department of Physics and Astronomy, Purdue University, 525 Northwestern Avenue, West Lafayette, IN 47907-2036, USA\\
$^{2}$Department of Physics, The George Washington University, Washington, DC 20052, USA\\
}

\date{Accepted XXX. Received YYY; in original form ZZZ}

\pubyear{2018}

\begin{document}
\label{firstpage}
\pagerange{\pageref{firstpage}--\pageref{lastpage}}
\maketitle

\begin{abstract}
The neutron star merger, GW170817, was followed by an optical-infrared transient (a kilonova) which indicated that a substantial ejection of mass at trans-relativistic velocities occurred during the merger. Modeling of the kilonova is able to constrain the kinetic energy of the ejecta and its characteristic velocity but, not the high-velocity distribution of the ejecta. Yet, this distribution contains crucial information on the merger dynamics. In this work, we assume a power-law distribution of the form $E(>\beta\Gamma)\propto(\beta\Gamma)^{-\alpha}$ for the energy of the kilonova ejecta and calculate the non-thermal signatures produced by the interaction of the ejecta with the ambient gas. We find that ejecta with minimum velocity $\beta_0\simeq 0.3$ and energy $E\sim 10^{51}$~erg, as inferred from kilonova modeling, has a detectable radio, and possibly X-ray, afterglow for a broad range of parameter space.
This afterglow component is expected to dominate the observed emission on a timescale of a few years post merger and peak around a decade later.
Its light curve can be used to determine  properties of the kilonova ejecta and in particular the ejecta velocity distribution $\alpha$, the minimum velocity $\beta_0$ and its total kinetic energy $E$. We also predict that an afterglow rebrightening, that is associated with the kilonova component, will be accompanied by a shift of the centroid of the radio source towards the initial position of the explosion.

\end{abstract}




\section{Introduction}

The gravitational wave source GW170817 marks the first merger of neutron stars ever detected (\citealt{abbottGW}). A variety of electromagnetic (EM) counterparts were detected following the GW170817 trigger (\citealt{Abbott2017multi}). These counterparts are powered by ejecta and outflows produced from the merger which range from relativistic to non-relativistic velocities. The relativistic outflows are associated with collimated jets, that produce the EM transient known as a gamma-ray burst (GRB). The trans-relativistic and non-relativistic ejecta consists of material that  has become unbound during the merger and outflows released by the remnant material following the merger (e.g., disc winds), which powers the transient known as a kilonova (KN; \citealt{li1998,kulkarni2005,rosswog2005,metzger2010}). 

Both a GRB and KN were detected in GW170817, which, for the first time, directly pointed their origin to NS mergers (e.g., \citealt{abbott2017gw+grb,pozanenko2018,goldstein2017,soares2017,coulter2017,hu2017,savchenko2017,utsumi2017}). The KN consisted of thermal optical/infrared emission which peaked a few days after the detection of GW170817 and is in general agreement with expectations from KN models (\citealt{metzger2010,barnes2013,tanaka2013,HBP2018}). 
The KN is believed to be produced by the radioactive decay of the heaviest elements synthesized in this ejecta. Modeling the multi-wavelength light curves of the KN, one can infer a characteristic ejecta velocity of $\sim 0.1c-0.3c$ and kinetic energy of $E\sim 10^{51}$~erg (e.g., \citealt{cowperthwaite2017,nicholl2017,arcavi2018}).  However, it is not clear how this energy is distributed within the ejecta, which could contain important information on the merger dynamics (e.g., \citealt{radice2018}).

The outflows from a NS merger drive an external shock as they propagate through the surrounding medium, this shock accelerates the particles in the external medium causing them to radiate primarily by synchrotron emission. This non-thermal emission associated with the shock is called the afterglow, we will refer to the afterglow associated with the jet as the ``GRB afterglow'' and the afterglow associated with the KN ejecta as the ``KN afterglow'' in this manuscript.

The GRB including its long term X-ray to radio afterglow that has been detected from GW170817 so far can be explained as originating from a structured jet with a narrow core (with an opening angle of $\sim 2^{\circ}-6^{\circ}$) which is misaligned by $\sim 20^{\circ}-30^{\circ}$ with respect to our line of site (e.g., \citealt{alexander2018,davanzo2018,kathir2017,kathir2018,lamb2017,lazzati2018,margutti2018,mooley2018superluminal,Ghirlanda2018,troja2018,xie2018,Beniamini2019}). Observation of the rather steep decline of this afterglow after its peak indicates the entire jet has come into view. The jet's true energy can, therefore, be constrained by the afterglow modeling and is inferred to be $\sim 10^{50}$~erg (e.g., \citealt{vaneerten2018}), not unlike that of other short-duration GRBs \citep{Fong2015}. 

In addition to the GRB afterglow, the KN will have its own afterglow. However, the KN ejecta is slower compared to the jet (but has comparable, if not more, energy), which will lead to the KN afterglow peaking at much later times compared to the GRB afterglow (\citealt{nakar2011,piran2013,alexander2017,hotokezaka2018,radice2018}). The KN afterglow contains important information on how the energy is distributed within the KN ejecta and its light curve will be sensitive to any velocity stratification within this ejecta. The KN afterglow is the focus of this paper, we will investigate how the KN afterglow differs for a fairly general energy distribution and develop an analytic framework which can be utilized to constrain properties of the KN ejecta using detections (or even non-detections) of the KN afterglow, with an application to GW170817 as an example.

The structure of the paper is the following. Sec. \ref{knmodel} describes our modeling of the KN ejecta, its interaction with the ambient gas, as well as the resulting synchrotron emission from these interactions. In Sec. \ref{gw170817app}, we apply the KN afterglow model to GW170817, making specific predictions on when this afterglow component could be observed and which ejecta properties can be probed by observations. Sec. \ref{conclusion} summarizes our conclusions.

\section{The KN blast wave and its afterglow}
\label{knmodel}

During the neutron star merger, a modest fraction of a solar mass is expected to be ejected. The total mass ejected and the angular and velocity distribution of the ejecta depend on the total mass of the progenitor system, the mass ratio of the neutron stars, and the nuclear equation of state (see, e.g., \citealt{baiotti2017review} for a review). In Sec. 2.1, we present a quite general parametrization of the ejecta velocity distribution and proceed to calculate the velocity profile of the shock driven by the ejecta into the ambient gas. Sec. 2.2 focuses on the synchrotron emission from electrons accelerated at this shock.     

\subsection{Properties of the kilonova ejecta and Dynamics of the blast wave}
\label{kndynamics}

About half a day after the GW170827 trigger, an optical counterpart, AT2017gfo, was discovered (\citealt{arcavi2017,lipunov2017,smartt2017,valenti2017}). The observed emission started out blue in color,  before rapidly evolving over the following days. Broad spectra at day $\sim$2.5 post trigger indicate the presence of distinct optical and near infrared emission components. Subsequently, the blue component fainted rapidly and the overall spectrum softened, peaking in the near-infrared. The widely accepted interpretation for this emission is the KN model (\citealt{cowperthwaite2017,chornock2017,nicholl2017,evans2017,Abbott2017multi,McCully2017,drout2017,tanvir2017}). The red KN is likely to be associated with slower ejecta ($\beta\sim 0.1$), while the blue KN can be powered by faster ($\beta\sim 0.3$) ejecta. The former component may have become unbound due to tidal interactions and is predominantly distributed along the equatorial plane while the latter is possibly associated with remnant disk outflows and shock ejected material primarily distributed away from the equatorial plane (\citealt{kasen2015,kasen2013,barnes2013,metzger2014,perego2017,radice2016,hotokezaka2013,siegel2018,fernandez2018}; see, however, \citealt{waxman2018,yu2018} for a different interpretation). 

Independent of the details and assumptions involved in the light-curve modeling, the post-merger ejecta appear to be fairly massive with total mass $\gtrsim 0.05M_\odot$ and velocity $\sim 0.1-0.3$~c, corresponding to a kinetic energy in excess of $\sim 10^{51}$ erg. These ejecta masses are broadly consistent with the estimated $r$-process production rate required to explain the heavy element abundances of the Universe, providing the first direct evidence that binary neutron star mergers can be a dominant site of $r$-process enrichment (\citealt{BHP2016,kasen2017,metzger2017review,siegel2017,BDS2018,rosswog2018,HBP2018}).

The modeling of the KN emission is mostly sensitive to the low-end of the velocity distribution of the ejecta\footnote{However, lanthanide-rich materials moving at slower velocities could also be missed in the optical, near infrared as the emission in those bands starts being dominated by the GRB afterglow.}, with which the bulk of the ejecta is moving. Modeling of this thermal transient does not give much information about any high-velocity tail of the ejecta. However, simulations studying NS mergers find that ejecta powering the KN are likely to be broadly distributed in energy as a function of $\beta\Gamma$, where $\Gamma$ is the Lorentz factor of the ejecta (e.g., \citealt{hotokezaka2018,fernandez2018,radice2018}). Motivated by these findings, we assume a power law distribution of the form $E(>\beta\Gamma)\propto(\beta\Gamma)^{-\alpha}$ for the energy of the KN blast wave, where  $\alpha$ typically varies between 3 to 5. The distribution is normalized to the total energy ($E$) at some minimum velocity ($\beta_0$). Guided by observations of GW170817, we can assume $E(>\beta_0\Gamma_0)=E=10^{51}$ erg and $\beta_0= 0.3\,(0.1) $ for the fast (slow) component.

The stratified ejecta expands driving a shock into the ambient gas.   
We assume a uniform external medium and approximate the total energy of the blast wave as $E\propto(\beta\Gamma)^2R^3$, with $R$ the radius of the blast wave. This expression encapsulates the dynamical evolution of the blast wave during the relativistic (\citealt{blandford1976}) and non-relativistic (\citealt{sedov1959}) phases. Given the assumed power law distribution for the ejecta's kinetic energy $E$, the blast velocity initially evolves with radius as $\beta\Gamma\propto R^{-\frac{3}{\alpha+2}}$, where the blast wave is continuously refreshed by slower, more energetic ejecta governed by the power law index $\alpha$. This continues until the total energy in the shocked external medium is comparable to that of the ejecta, after which point the total energy of the blast wave remains constant (assuming an adiabatic evolution) and the blast velocity evolves as $\beta\Gamma\propto R^{-\frac{3}{2}}$. The radius at which this transition occurs will be called the ``deceleration'' radius and is given by $R_{\rm dec}=(3E_{\rm iso}/4\pi \Gamma_0(\Gamma_0-1)nm_{\rm p}c^2)^{\frac{1}{3}}$, where  $m_p$ is the proton mass, $c$ is the speed of light and $E_{\rm iso}$ is the isotropic equivalent energy of the blast wave. The $R_{\rm dec}$ corresponds to the deceleration radius of the slowest material (which also carries the majority of the energy). 

The isotropic equivalent energy $E_{\rm iso}$ of each component is related to the true energy ($E$) by $E=f_{\Omega}E_{\rm iso}=\frac{1}{2}E_{\rm iso}\,\int{\sin}\theta {\rm d}\theta$, with $f_{\Omega}$ the solid angle fraction of the blast wave. Employing spherical coordinates, we assume the ejecta is distributed uniformly in the azimuthal direction from $\phi=0$ to $2\pi$, and divide the polar extent of the blast wave into two main components. Since the slower, ``red'' component is likely to be distributed on the equatorial plane, we assume $\theta=60^{\circ}-120^{\circ}$ for its polar extent. The fast component (associated with the ``blue'' KN) is assumed to be distributed within $\theta=15^{\circ}-60^{\circ}$ for one hemisphere and  $\theta=120^{\circ}-165^{\circ}$ for the other hemisphere. Changing the angular extent and viewing angle will not considerably affect our results since bulk of the KN ejecta is not relativistic, making beaming effects modest (see last paragraph of Sec. \ref{knmodelling} for further details).

\subsection{Modeling the KN afterglow}
\label{knmodelling}

The KN blast wave drives a shock through the external medium, energizing the swept up particles and causing them to radiate via synchrotron emission. This emission is termed the afterglow. We assume that the electrons are accelerated into a power-law distribution above a minimum Lorentz factor $\gamma>\gamma_{\rm m}$. The analytic expressions for the radiated flux used or derived below are applicable only if the observing frequency is between the minimum ($\nu_{\rm m}$) and cooling ($\nu_{\rm c}$) frequencies. The plotted light curves, however (which are calculated semi-analytically), include the effects of synchrotron and Compton cooling of the electrons as well as emission during the ``deep-Newtonian" phase, applicable when the the minimum Lorentz factor of the electrons $\gamma_{\rm m}\sim 1$  (\citealt{sironi2013}). 

The KN afterglow emission peaks at the deceleration radius, and the peak flux density can be expressed as (e.g., \citealt{nakar2011})
\beq
F_{\rm \nu, pk}\approx (115\,{\rm \mu Jy})\,\epsilon_{\rm e,-1}^{p-1}\,\epsilon_{\rm B,-3}^{\frac{p+1}{4}}\,n_{-2}^{\frac{p+1}{4}}\,\beta_{0}^{\frac{5p-7}{2}}\,E_{51}\,\nu_{9.5}^{\frac{1-p}{2}}\,d_{26}^{-2},
\label{fpeak}
\eeq
where the prefactor is determined for $p=2.2$, but does not change by more than a factor 3 when $p$ is varied from 2.1--2.5. Here $\epsilon_{\rm e}$ and $\epsilon_{\rm B}$ are the fractions of the total energy in the shocked electrons and magnetic fields of the shocked fluid respectively, $n$ is the number density of the uniform external medium, $p$ is the power law slope of the distribution of shocked electrons, $E$ is the true energy (integrated over velocity) of the KN blast wave, $\nu$ is the observing frequency and $d$ is the distance to the source. All quantities are in cgs units and we use the notation $Q_x=Q/10^x$.

The time of peak can be obtained by relating the observer time to the radius of the blast wave ($R$), and substituting $R=R_{\rm dec}$ to obtain $t_{\rm dec}\approx\int^{R_{\rm dec}}_0\frac{dr}{\beta(r)c}(1-\beta(r))$. Here we assume the observer is within line of sight of the KN blast wave. The minimum velocity ($\beta_0$) is expected to be, at most, mildly relativistic, in this we can assume $\beta_0\lesssim 0.5$, and obtain an analytic approximation for the observed peak time as  
\beq
t_{\rm dec}=t_{\rm pk}\approx (3.3 {\rm yr})\left(\frac{E_{\rm iso, 51}}{n_{-2}}\right)^{\frac{1}{3}}\, \beta_{0}^{-\frac{2}{3}}\left(\frac{2+\alpha}{\beta_0(5+\alpha)}-1\right).
\label{tpeakgeneral}
\eeq
As mentioned in Sec. 2.1, $\alpha$ typically varies between $3-5$ and will be closely examined in this work. For comparison, we will also consider an extreme case $\alpha\rightarrow\infty$, which corresponds to a single velocity component for the blast wave.
For the cases where $\alpha$ is between 3 and 5, one finds that the peak times vary by less than a factor $\sim 1.5$, hence for these cases, we can fix the value of $\alpha$ (here we will use $\alpha=4$). Then the peak time can be well approximated by a much simpler form
\beq
t_{\rm pk}(3\lesssim\alpha\lesssim5)\approx(8.5 {\rm yr})\left(\frac{E_{\rm iso, 51}}{n_{-2}}\right)^{\frac{1}{3}}\,\beta_{0,-0.5}^{-\frac{13}{6}}.
\label{tpeak1}
\eeq
When $\alpha\rightarrow\infty$ (corresponding to a single velocity component) the peak time can be approximated as
\beq
t_{\rm pk}(\alpha\rightarrow\infty)\approx(22.5 {\rm yr})\left(\frac{E_{\rm iso, 51}}{n_{-2}}\right)^{\frac{1}{3}}\, \beta_{0,-0.5}^{-\frac{5}{3}}.
\label{tpeak2}
\eeq

Before the peak (especially at early times of $\sim 1$ yr), the blast wave can be mildly relativistic. For these situations the slope of the light curve can be expressed as (\citealt{barniolduran2015,kathir2016})
\beq
s=\frac{3\alpha-6(p-1)}{8+\alpha},
\label{slope}
\eeq
where for $\alpha\rightarrow\infty$, we obtain the expected $t^3$ rise for associated with a spherical, single velocity component blast wave. Hence, the KN afterglow light curve (before peak) can be expressed as a function of time as
\beq
F_{\rm\nu, KN}\,(t)=F_{\rm \nu,pk}\left(\frac{t}{t_{\rm p}}\right)^s.
\label{KNflux}
\eeq

We now have three observables which can be used to constrain the properties of the KN, the peak flux ($F_{\rm \nu,pk}$), peak time ($t_{\rm pk}$) and slope ($s$).

\begin{figure}
\includegraphics[width=\columnwidth]{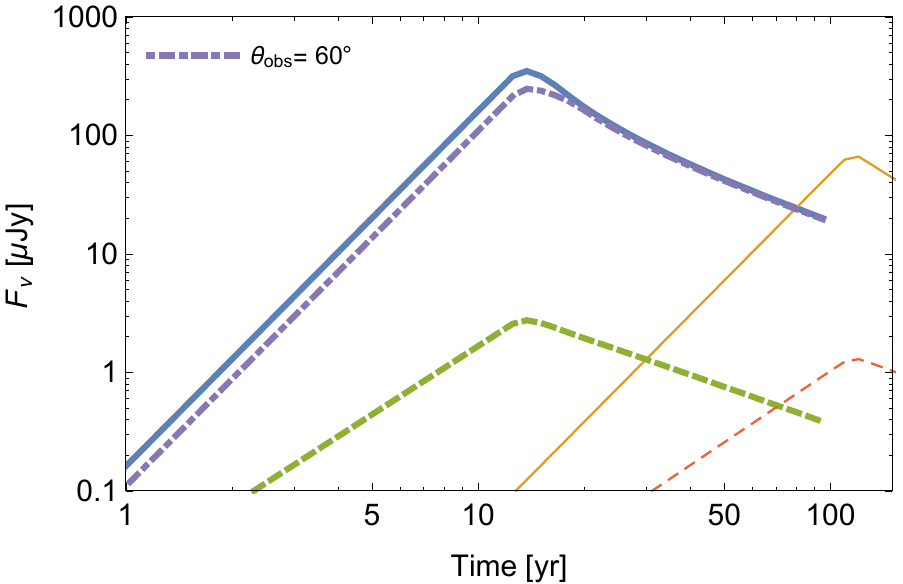}
\caption{ KN afterglow light curves for the fast ($\beta_0=0.3$, thick lines) and slow ($\beta_0=0.1$, thin lines) components in radio (3GHz, solid lines) and X-ray (1 keV, dashed lines) wavelengths. The magnitude of the X-ray flux density has been multiplied by $10^4$. The parameters used are $\alpha\rightarrow\infty,\,E=10^{51}\,{\rm erg},\epsilon_{\rm e}=0.1,$ $\epsilon_{\rm B}=10^{-3},$ $\theta_{\rm obs}=30^{\circ}$ (except for the dot-dashed line), $n=0.1\,{\rm cm^{-3}}$. The density is on the higher end compared to typical afterglow models of GW170817 to give a best case scenario for the detectability of the KN afterglow. It is evident that even for this best case, the rise and peak of the slow component occurs much later and is fainter compared to that of the fast component. Therefore, in this work we will only focus on the afterglow of the fast component, which is relevant for the current timescale of GW170817. Dot-dashed line shows radio afterglow of fast component for $\theta_{\rm obs}=60^{\circ}$ and is almost identical to the $\theta_{\rm obs}=30^{\circ}$ light curve, demonstrating that viewing angle effects are not significant for the KN afterglow.}
    \label{KNcomparison}
\end{figure}

Fig. \ref{KNcomparison} shows afterglow light curves for the fast (associated with the ``blue'' KN) and slow (associated with the ``red'' KN) components having $\beta_0=0.3\,$ {and} $\, 0.1$ respectively, in a uniform external density $n=0.1$ cm$^{-3}$, for an observing angle $\theta_{\rm obs}=30^{\circ}$ (except for the dot-dashed line). 
The KN afterglow light curves are produced semi-analytically following the same method as in (\citealt{kathir2018}). Where analytic expressions for the synchrotron emission in a forward shock (e.g., \citealt{sari1998,granot2002}) are used to find the radiated flux in the co-moving frame of the shock, transforming this flux to the observer frame and summing the flux over the entire blast wave while taking into account differences in photon arrival time. These calculations also take into account cooling of the electrons due to synchrotron and self-synchrotron Compton energy losses (e.g., \citealt{beniamini2015}), and emission during the deep-Newtonian phase (\citealt{sironi2013}), with modifications to include an energy distribution in the blast wave as described in Sec. \ref{kndynamics}. Synchrotron self-absorption is not considered in these calculations, which is justifiable since we have verified that the turn-over frequency always lies below the observed bands. This is because the emitting region at $\sim 1$~yr after the burst is very extended and thus remains optically thin down to very low frequencies.

The afterglow of the slow component (thin lines in Fig. \ref{KNcomparison}) peaks much later ($\sim 100$ yrs) compared to the fast component ($\sim 10$ yrs). Therefore, focusing on the current timescales of $\sim 1$ yr since GW170817, we will restrict our discussion to the afterglow of the fast component throughout this manuscript. The dot-dashed line in Fig. \ref{KNcomparison} shows the radio light curve of the fast component for $\theta_{\rm obs}=60^{\circ}$, with all other parameters kept the same as before. This light curve is almost identical to the $\theta_{\rm obs}=30^{\circ}$ light curve, which demonstrates that the exact angular geometry and its affect on beaming are of minor importance here (as mentioned in Sec \ref{kndynamics}).

\section{Application to GW170817}
\label{gw170817app}

In this section we apply our KN afterglow model to GW170817 as an example case, with focus on how properties of the ejecta can be constrained using detections (or even a non-detection) of the KN afterglow emission. In Sec~\ref{gw170817kn}, we present general expressions with the prediction of the afterglow flattening/rebrigheting associated with the emergence of the KN afterglow, and summarize what we can learn from this emergence. Sec.~\ref{3.2} focuses on the much smaller parameter space of the model for which one can fit the observed GRB afterglow with a structured jet model. In this case, our predictions for the expected emission from KN component become much more definite.

\subsection{Constraining the KN of GW170817}
\label{gw170817kn}

In order to produce example light curves for the KN afterglow, we have to assume some values for the microphysical parameters ($\epsilon_{\rm e}$, $\epsilon_{\rm B}$, particle index ($p$)), and the external density ($n$).  As a guide for our choice, we use typical parameters inferred from the fitting and observations of GRB170817A afterglow (associated with the jet), as well as typical parameters inferred for other afterglows of short GRBs, which are, $\epsilon_{e}=0.1,\,\epsilon_{\rm B}=10^{-3},\,n=10^{-2}\,{\rm cm}^{-3},\,p=2.2$  \citep{Nava2014,Santana2014,GvdH2014,beniamini2015,Fong2015,Beniamini2016,Zhang2015,BvdH2017,margutti2018,VanEerten2018review}.

\begin{figure}
\includegraphics[width=\columnwidth]{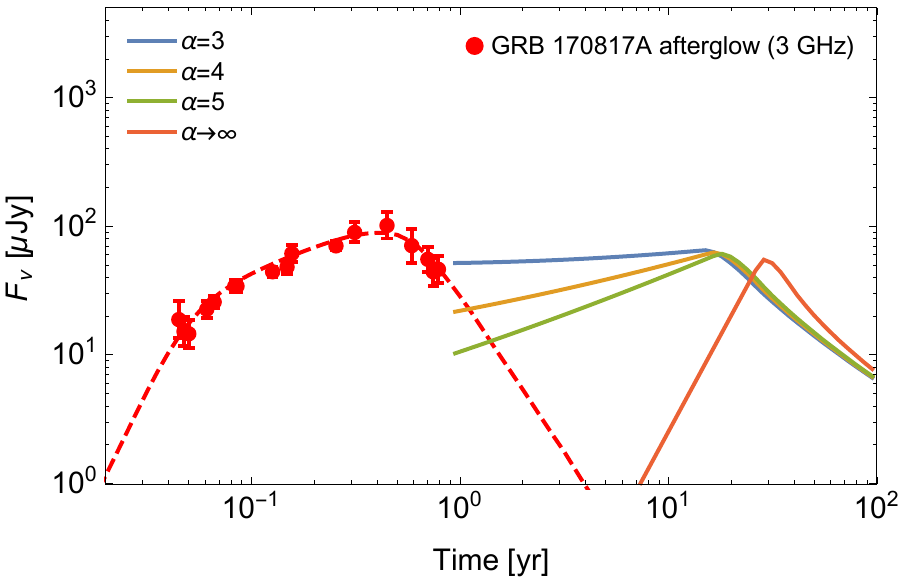}\\
\includegraphics[width=\columnwidth]{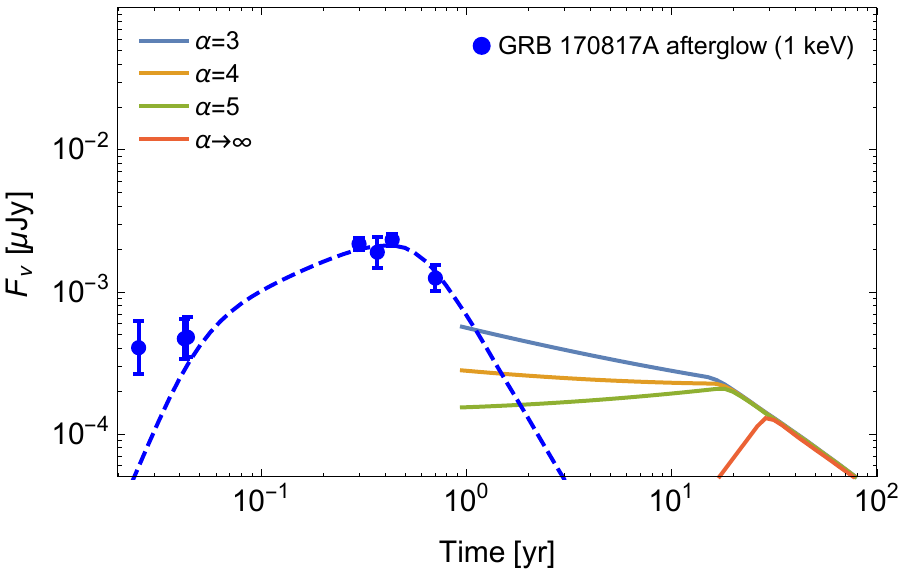}
\caption{A predicted rebrightening in the GW170817 afterglow. Observed data of GRB170817A afterglow (points) along with the afterglow model from \citep{kathir2018} (dashed lines) in radio (top panel) and X-ray (bottom panel). Solid lines show KN afterglow light curves for varying $\alpha$ (see insert labels), which peak at later times resulting in a flattening/rebrightening in the overall afterglow. The X-ray afterglow of the KN is above the cooling frequency for the times considered here. Therefore, detection in the radio is more favourable. Data points obtained from \citet{hallinan2017,alexander2018,margutti2018,mooley2018jet}.}
    \label{KNafterglow}
\end{figure}

Fig. \ref{KNafterglow} shows KN afterglow of the fast component in radio (3 GHz) and X-ray (1 keV) for different values of $\alpha$  with the ``typical'' parameters mentioned above. In the same plot we show observed data points for the afterglow of GRB170817A (data obtained from \citealt{hallinan2017,alexander2018,margutti2018,mooley2018jet}) along with the theoretical prediction of the afterglow from the structured jet model presented in \citet{kathir2018}. The KN afterglow light curves are produced semi-analytically following the same method as detailed in Sec. \ref{knmodel}.

Equation $\ref{KNflux}$ accurately reproduces the radio afterglow shown in the top panel of Fig. \ref{KNafterglow} before peak. However, this equation does not apply to the X-ray afterglow (bottom panel) since at 1 keV, the spectrum is above the cooling frequency for the timescales shown and parameters chosen. It might be more difficult to detect the KN afterglow in the X-ray band if it is in the fast cooling regime, therefore, we will focus our analysis on the radio emission for the remainder of this work. An X-ray detection of the KN afterglow, for more favorable parameters than our reference case, is however possible.

Fig. \ref{KNafterglow} and equation \ref{slope} show that observations of the rise in the afterglow can be used to constrain $\alpha$ well before the peak (provided $p$ is known, e.g., from multi-frequency observations). The peak time and flux can be used to constrain quantities such as $\beta_0$ and $E$. However, as we show below, even a non-detection of the KN afterglow can be used to constrain properties of the KN outflow.

\begin{figure}
\includegraphics[width=\columnwidth]{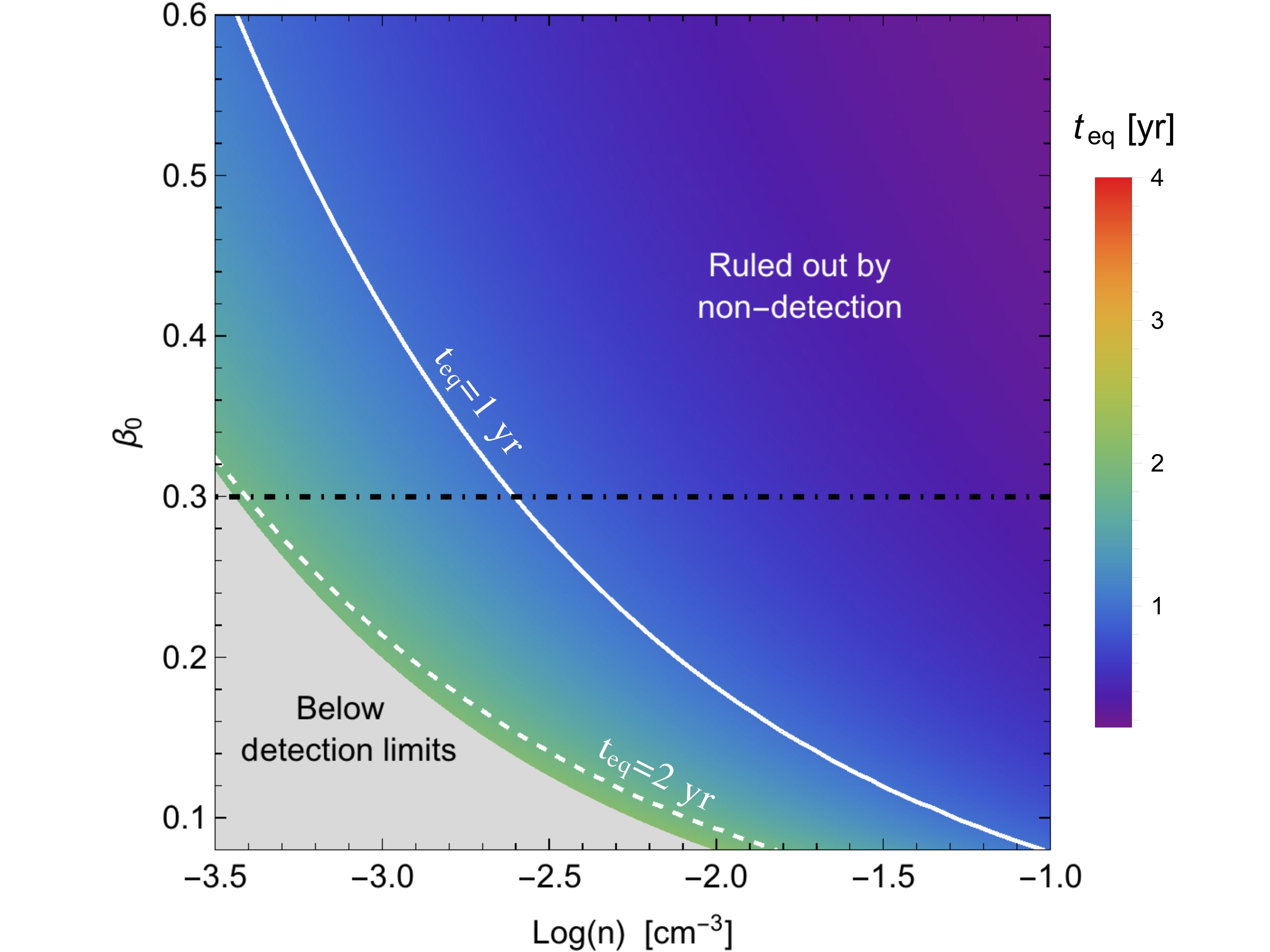}\\
\includegraphics[width=\columnwidth]{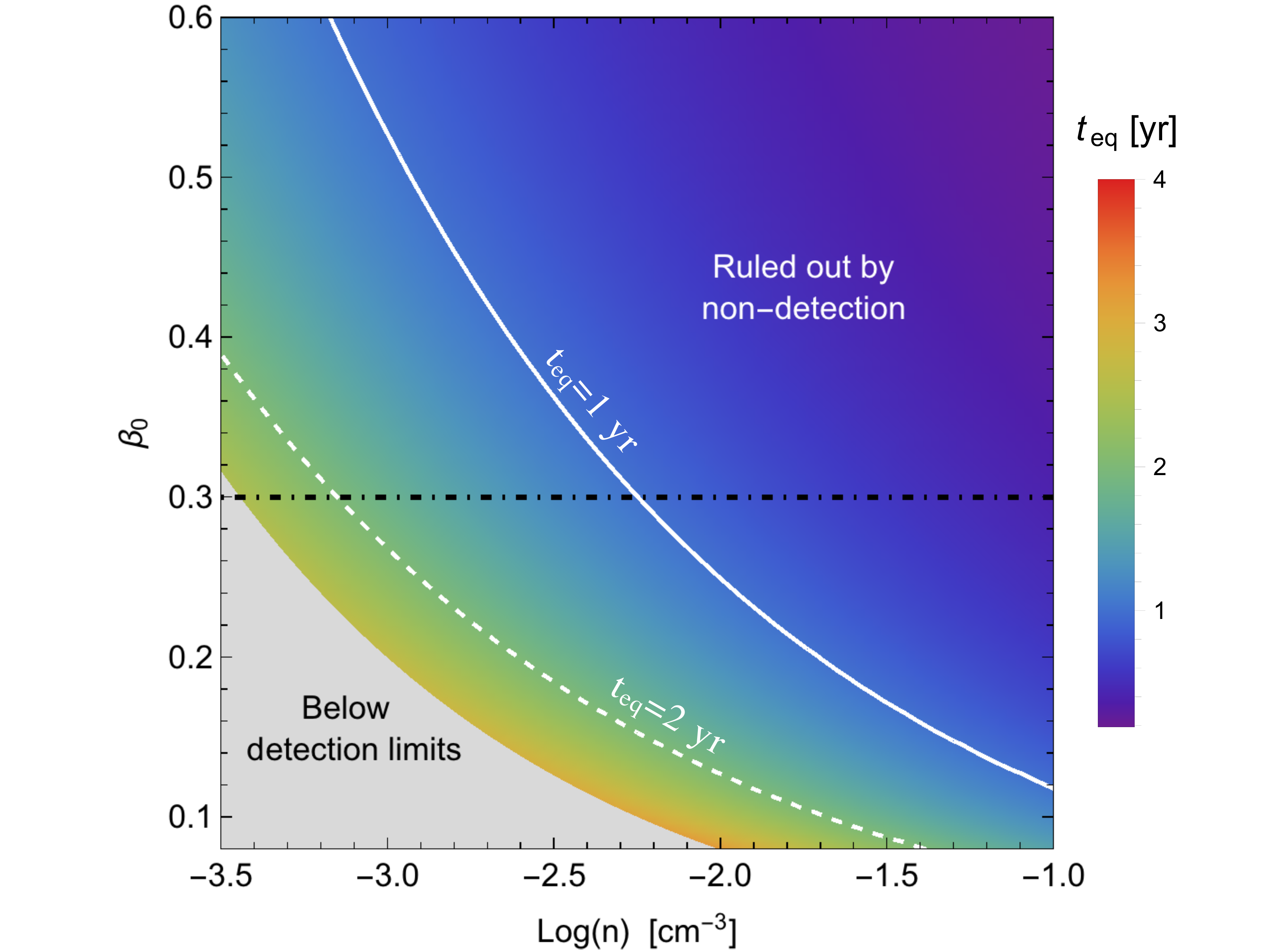}\\
\includegraphics[width=\columnwidth]{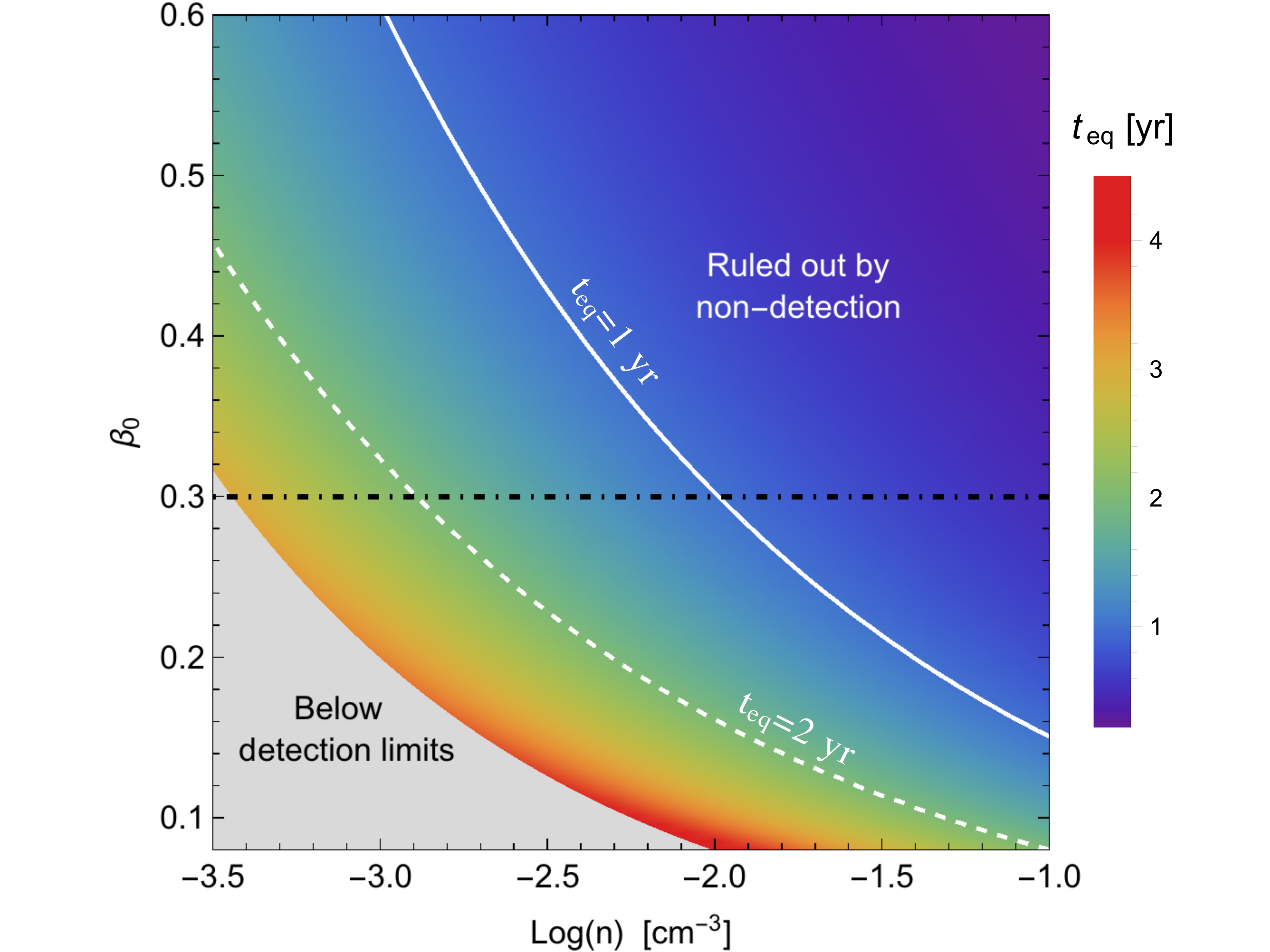}
\caption{Contour plot of the time of emergence of the KN afterglow $t_{\rm eq}$ (equation \ref{teq}) for minimum speed of the KN blast wave ($\beta_0$) vs. external density ($n$) fixing $\alpha=3$ (top panel), $\alpha=4$ (middle panel) and $\alpha=5$ (bottom panel). The solid and dashed lines show the $t_{\rm eq}=1$~yr and $t_{\rm eq}=2$~yr contours respectively. The peak of the KN afterglow in the gray, shaded region lies below the detectability limit ($F_{\rm \nu, p}\approx 5\,\mu$Jy) in radio and therefore is not detectable. Horizontal, dot-dashed line marks $\beta_0=0.3$ which is the typical velocity of the fast component inferred from observations of the blue KN. The emergence of the KN afterglow has not been detected yet implying $t_{\rm eq}\gtrsim 1$ yr, which corresponds to regions below the $t_{\rm eq}=1$ contour in the above figures.}
    \label{teqconstraints1}
\end{figure}

\begin{figure}
\includegraphics[width=\columnwidth]{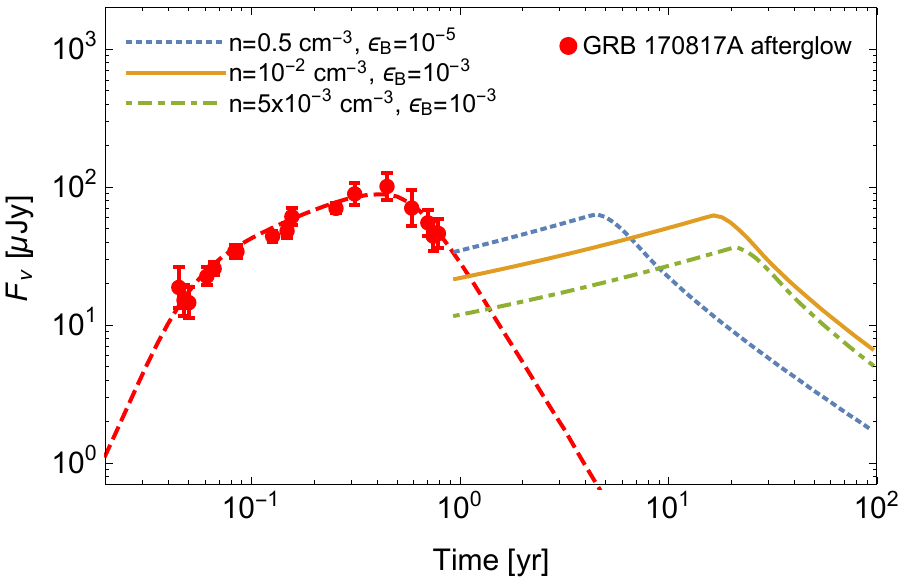}\\
\includegraphics[width=\columnwidth]{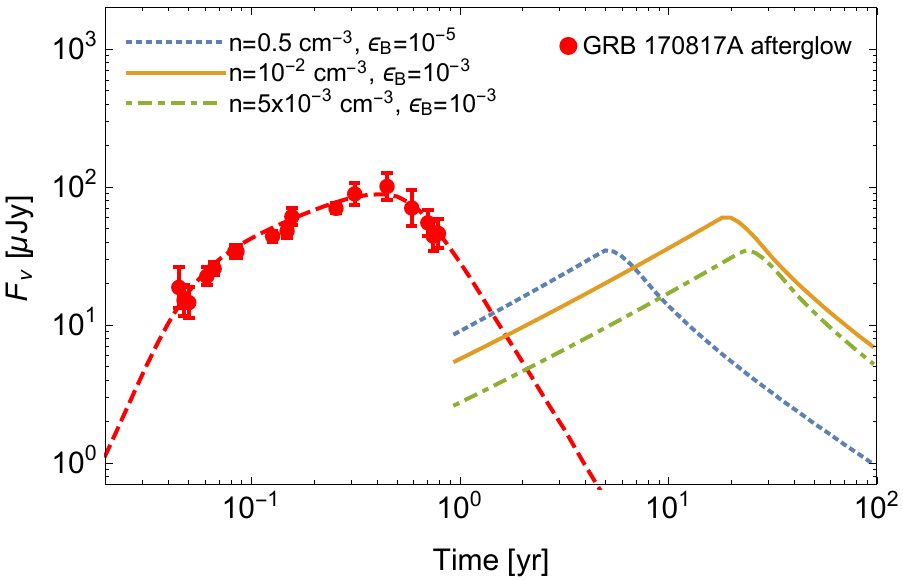}
\caption{Radio (3 GHz) data of GRB170817A afterglow (points) along with structured jet afterglow model from \citet{kathir2018} (dashed line) and KN afterglow light curves with fixed $\alpha=4$ (top panel) and $\alpha=6$ (bottom panel) for a range of $n$ and $\epsilon_{\rm B}$ that are typically inferred from fitting the GRB170817A afterglow. The solid line ($n=10^{-2}$ cm$^{-3}$, $\epsilon_{\rm B}=10^{-3}$) uses the same parameters as in our calculations in Sec~\ref{gw170817kn} . It is evident that varying the parameters within this range does not significantly alter the light curves. An afterglow rebrightening may be expected within $\sim$2 years after the merger.}
    \label{vary}
\end{figure}

We have observed the peak in the afterglow of GRB170817A at $\sim 150$ days, and the decline post-peak follows a power law in time with slope $\sim -2.4$ (\citealt{alexander2018,mooley2018jet}). Therefore, the decline of the GRB afterglow can be modelled as 
\beq
F_{\rm d}\,(t)\approx (100\, {\rm \mu Jy})\left(\frac{t}{0.44\,{\rm yr}}\right)^{-2.4}.
\label{AGflux}
\eeq

By equating \ref{KNflux} and \ref{AGflux}, we can find the time at which the flux from the afterglow and KN will be equal (\teq). After this time, it will be possible to detect the KN afterglow. The general expression for $t_{\rm eq}$ is 
\beq
t_{\rm eq}=\left(\frac{(100\mu{\rm Jy})(0.44 {\rm yr})^{2.4}\, t_{\rm p}^{s}}{F_{\rm \nu,p}}\right)^{\frac{1}{s+2.4}} = \left(\frac{14\, t_{\rm p}^{s}}{F_{\rm \nu,p}}\right)^{\frac{1}{s+2.4}}.
\label{teqgeneral}
\eeq
For $3\lesssim\alpha\lesssim5$, we can substitute expression \ref{tpeak1} for $t_{\rm p}$ and obtain
\beq
t_{\rm eq}\approx\left(\frac{0.2\,(8.3)^{\frac{3 \alpha -6 (p-1)}{\alpha +8}} \nu _{9.5}^{\frac{p-1}{2}} \text{$\beta $}_{0,-0.5}^{-\frac{6 (\alpha -5)+(5 \alpha +14) p}{2 (\alpha +8)}}}{f_{\Omega}\,E_{\text{iso},51}^{\frac{2(p+3)}{\alpha +8}}\,\epsilon _{B,-3}^{\frac{p+1}{4}}\,\epsilon _{e,-1}^{p-1}\, n_{-2}^{-\frac{\alpha(p+5)+16}{4 (\alpha +8)}}}\right)^{\frac{5 (\alpha +8)}{126-30p+27\alpha}},
\label{teq}
\eeq
where we have substituted $d=40$ Mpc for the distance to the source. If we substitute $\alpha=4$ and $p=2.2$ for example, $t_{\rm eq}$ becomes
\beq
t_{\rm eq}\approx(2 {\rm\,yr})\,n_{-2}^{-0.33}\,\beta_{0,-0.5}\,E_{\rm iso, 51}^{-0.31}\,\nu_{9.5}^{0.21}\,\epsilon_{\rm e,-1}^{-0.43}\,\epsilon_{\rm B,-3}^{-0.29}\,f_{\Omega}^{-0.36}.
\eeq
Fig. \ref{teqconstraints1} shows a contour plot of $t_{\rm eq}$ for $\beta_{0}$ vs $n$ where each panel corresponds to a different value of $\alpha$. The rest of the parameters used are $\epsilon_{\rm e}=0.1,\,\epsilon_{\rm B}=10^{-3},\, E_{\rm iso}=10^{51} \,{\rm erg},\,  p=2.2\,$ and $\nu=3$ GHz.

The rebrightening time $t_{\rm eq}$ can also be calculated for $\alpha\rightarrow\infty$ case using equation \ref{tpeak2}. We do not  focus on this case here since the afterglow at $t_{\rm eq}$ will most likely be too faint to detect (see Fig. \ref{KNafterglow}). However, the peak of the KN afterglow may still be bright enough to be detected. Interestingly, therefore, if the KN ejecta are characterized by a very steep velocity profile, its afterglow emission could fall below detectability limits in the near future, only to re-emerge several years, or even a decade, later.

We have not yet observed the emergence of the KN afterglow in GW170817, indicating that $t_{\rm eq}$ must be greater than the current observing time ($t_{\rm eq}\gtrsim 1$ yr). Using this condition on $t_{\rm eq}$, we can place some constraints on the dynamical and micro-physical quantities related to the KN afterglow (such as $\beta_0$ and $\alpha$) even without a detection. For example, if we substitute $\epsilon_{\rm e}=0.1,\, E_{\rm iso}=10^{51} \,{\rm erg},\,  p=2.2,\, \alpha=4$ and $\nu=3$ GHz, the condition $t_{\rm eq}\gtrsim 1$ yr yields
\beq
\beta_0\lesssim0.25\, n_{-2}^{-0.33}\,\epsilon_{\rm B,-3}^{0.28}.
\label{teqcond1}
\eeq
This equality ($t_{\rm eq}=1$ yr) is shown by the solid, white lines in Fig \ref{teqconstraints1} (with $\epsilon_{\rm B}=10^{-3}$). For regions above the $t_{\rm eq}=1$ yr contour, $t_{\rm eq}$ is less than 1 yr, which means these regions are excluded since no rebrightening/flattening has been observed in the afterglow of GW170817. For comparison, Fig. \ref{teqconstraints1} also shows the $t_{\rm eq}=2$ yr contour. In order for the KN afterglow to be detectable, at the very least, its peak flux must be greater than sensitivity limits of detectors. For radio observations, we will use a sensitivity limit $5\, \mu$Jy so that the detectability condition is $F_{\rm 
\nu, p}\gtrsim 5\,\mu$Jy, with $F_{\rm 
\nu, p}$ given in equation \ref{fpeak} . Substituting the same parameters used to obtain \ref{teqcond1}, this detectability condition yields
\beq
\beta_{0}\gtrsim0.08\,n_{-2}^{-0.4}\,\epsilon_{\rm B,-3}^{-0.4}.
\label{senscondition}
\eeq
Regions which do not satisfy this condition are shaded gray in Fig. \ref{teqconstraints1} (using $\epsilon_{\rm B}=10^{-3}$) and the KN afterglow will not be detectable for parameters in this region. The horizontal dot-dashed line marks where $\beta_0=0.3$, which is the characteristic velocity inferred from observations of the blue KN (see Sec. \ref{knmodel}). The range of densities for the x-axis of Fig. \ref{teqconstraints1} was chosen guided by afterglow modelings of GW170817 (\citealt{lazzati2018,vaneerten2018,resmi2018,wu2018}). From these conditions, we can begin to constrain properties of the KN. For example, from Fig. \ref{teqconstraints1}, we see that for $\alpha=3$ (top panel), if $\beta_0= 0.3$, the external density must be $\lesssim 0.005$ cm$^{-3}$, otherwise the KN afterglow would have been detected by now.

\subsection{Combined Constraints from the jet afterglow}
\label{3.2}

 Using the analytic expressions given in Sec. \ref{knmodel} the analysis carried out in Sec. \ref{gw170817app} can be done for different values of microphysical parameters ($\epsilon_{\rm e}$ and $\epsilon_{\rm B}$) and external density ($n$), provided that the observed frequency lies between $\nu_{\rm m}$ and $\nu_{\rm c}$ for the choice of parameters. In this example (Fig. \ref{KNafterglow}), values that provided a good fit for the afterglow observations of GRB170817A were used.
The typical values of density and $\epsilon_{\rm B}$ found from fitting the GRB afterglow of GW170817 range from $\sim 10^{-3}-10^{-1}$ cm$^{-3}$ and $\sim 10^{-5}-10^{-3}$ respectively (e.g., \citealt{resmi2018,vaneerten2018,wu2018}), where higher densities require lower values of $\epsilon_{\rm B}$ to fit the afterglow (see e.g., Fig. 3 of \citealt{wu2018}). This inverse proportionality becomes evident if one uses the expression for the peak flux of a GRB afterglow (e.g., \citealt{nakar2002}) and expresses $n$ in terms of $\epsilon_{\rm B}$ (keeping other parameters fixed) to find $n\propto\epsilon_{\rm B}^{-1}$ (this is also the case for the peak of the KN afterglow; equation \ref{fpeak}). 

Fig. \ref{vary} shows the radio afterglow light curves of the KN for these range of densities and $\epsilon_{\rm B}$ to demonstrate how the light curves would vary for different choices of these parameters which provide reasonable fits for the GRB afterglow. 
Fig. \ref{vary} shows that within this range (that spans 2 orders of magnitude), the flux density of the afterglow does not change by more than a factor $\sim$ 2 when compared to our example case of $n=10^{-2}$ cm$^{-3}$ and $\epsilon_{\rm B}=10^{-3}$ (see Fig. \ref{vary}). Which means our conclusions will remain the same for this broad range of parameter space obtained when fitting the GRB afterglow.
An afterglow rebrightening, due to the emergence of the KN component, may be expected within $\sim$2 years after the merger.

\begin{figure}
\includegraphics[width=\columnwidth]{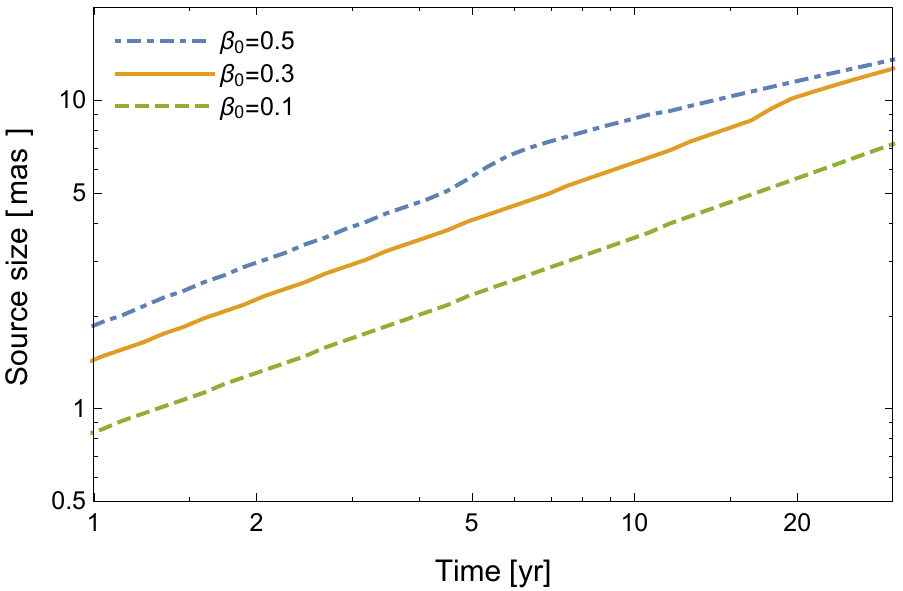}\\
\includegraphics[width=\columnwidth]{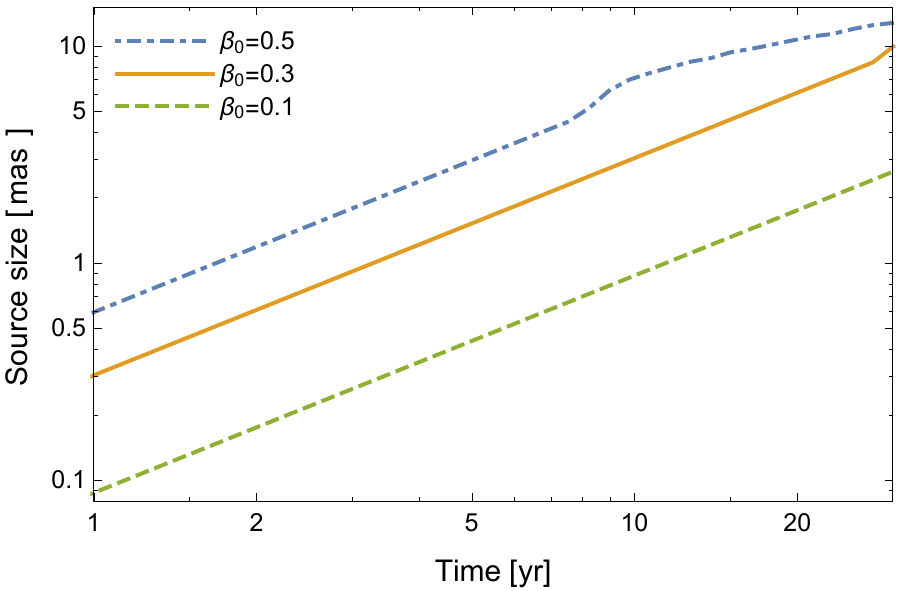}
\caption{Angular size of the source (associated with KN afterglow) vs. time for the case where $\alpha=4$ (top panel) and $\alpha\rightarrow\infty$ (bottom panel) fixing $n=10^{-2}$ cm$^{-3}$ and $E=10^{51}$ erg. Three different values of $\beta_0$ are shown in each case. Given a typical size of the source of $\sim 10$~mas at around $\sim$1 decade post trigger and corresponding flux density $\sim 50-100$~$\mu$Jy, resolving the source may be possible if the KN rebrightening is observed.}
    \label{sourcesize}
\end{figure}
\section{Discussion/Conclusions}
\label{conclusion}

GW170817 was followed by an optical-IR transient, AT2017gfo, which revealed that the merger was accompanied by a substantial ejection of mass $M_{\rm ej}\sim 0.05M_{\odot}$ at trans-relativistic velocities $\beta_0\sim 0.1-0.3$. The KN modeling is able to constrain the kinetic energy of the ejecta and its characteristic velocity but is less sensitive to the high-velocity distribution of the ejecta. Yet, this distribution contains crucial information on the merger dynamics. In this work, we assume a power-law distribution of the form $E(>\beta\Gamma)\propto(\beta\Gamma)^{-\alpha}$ for the energy of the KN ejecta and calculate the resulting afterglow powered by the KN ejecta. We find that: 
\begin{enumerate}
\item A fast KN component with minimum velocity $\beta_0\simeq 0.3$ and energy $E\sim 10^{51}$~erg, which is likely responsible for the observed blue KN emission, can produce a detectable radio, and possibly X-ray, afterglow for a broad range of the parameter space.
\item For $3\lesssim\alpha\lesssim 6$, the KN afterglow is expected to emerge, by dominating the afterglow emission, on a timescale of a few years and peak around a decade later. 
\item For steep values of $\alpha\to \infty$, the afterglow emission can drop below detectability levels before the KN afterglow emerges on a decade time scale.  
\item The time of emergence $t_{\rm eq}$ (equation \ref{teqgeneral}), the rise slope of the light curve $s$ (equation \ref{slope}) and the peak time $t_{\rm p}$ (equation \ref{tpeakgeneral}) can be used to determine  properties of the KN ejecta, in particular, the ejecta velocity distribution $\alpha$, the minimum velocity $\beta_0$ and its kinetic energy $E$.
\end{enumerate}

The GRB afterglow of GW170817 has been characterized by a superluminal apparent speed in the radio (\citealt{mooley2018superluminal}). This finding is consistent with the misaligned jet interpretation used to describe the non-thermal emission observed so far from this event (see also \citealt{zrake2018} for an investigation into the radio map of GW170817). On the other hand, the KN blast is expected to quasi isotropic with its radio image centered around the merger location.  We, therefore, predict that any afterglow rebrightening --marking the emergence of the KN component-- will be accompanied by a shift of the centroid of the radio towards the initial position of the explosion. Ultimately, the KN afterglow may be sufficiently bright and extended for the source to be resolved, Fig.~\ref{sourcesize} shows the size of the source vs. time with $n=10^{-2}$ cm$^{-3}$ and $E=10^{51}$ erg. The size is obtained by calculating the extent of the blast wave which contributes to half of the afterglow emission and projecting this extent perpendicular to the line of sight. For $\beta_0\simeq 0.3$, the typical size of the source is $\sim 10$~mas at around twenty years post merger.

As of the time of writing of this work, the afterglow emission from GW170817 is declining quite steeply (\citealt{mooley2018jet}), which is expected from a structured jet model that is moderately misaligned with respect to our line of sight (e.g., \citealt{KG2003,Salmonson2003,kathir2018,lamb2018,vaneerten2018,BN2019}). We find here that the ejecta responsible for the KN may cause the afterglow light curve to rebrighten in the near future and be detectable for decades to come. 

\section*{Acknowledgements}

AK acknowledges support from the Bilsland Dissertation Fellowship. AK and DG acknowledge support from NASA grant NNX17AG21G and the NSF grant number 1816136. This research was supported in part through computational resources provided by Information Technology at Purdue, West Lafayette, Indiana.





\bibliographystyle{mnras}
\bibliography{references}




\bsp	
\label{lastpage}
\end{document}